\documentstyle[12pt]{article}
\newcommand{\newsection}[1]{
\vspace{15mm}
\pagebreak[3]
\addtocounter{section}{1}
\setcounter{equation}{0}
\setcounter{subsection}{0}
\setcounter{footnote}{0}
\addcontentsline{toc}{section}{\protect
\numberline{\arabic{section}}{{\rm #1}}}
\begin{center}
{\large \sc \thesection. #1}
\end{center}
\nopagebreak
\medskip
\nopagebreak}
\newcommand{\newsubsection}[1]{
\vspace{1cm}
\pagebreak[3]
\addtocounter{subsection}{1}
\addcontentsline{toc}{subsection}{\protect
\numberline{\arabic{section}.\arabic{subsection}}{#1}}
\noindent{ \sc  #1}                 
\nopagebreak
\vspace{2mm}
\nopagebreak}
\newlength{\extraspace}
\newlength{\extraspaces}
\newcommand{\be}{\begin{equation}
\addtolength{\abovedisplayskip}{\extraspaces}
\addtolength{\belowdisplayskip}{\extraspaces}
\addtolength{\abovedisplayshortskip}{\extraspace}
\addtolength{\belowdisplayshortskip}{\extraspace}}
\newcommand{\ee}{\end{equation}}
\newcommand{\ba}{\begin{eqnarray}
\addtolength{\abovedisplayskip}{\extraspaces}
\addtolength{\belowdisplayskip}{\extraspaces}
\addtolength{\abovedisplayshortskip}{\extraspace}
\addtolength{\belowdisplayshortskip}{\extraspace}}
\newcommand{\ea}{\end{eqnarray}}
\newcommand{\nonu}{\nonumber \\[2mm]}
\newcommand{\is}{& \!\! = \!\! &}
\newcommand{\ie}{{\it i.e.\ }}

\newcommand{\cP}{{P}}
\newcommand{\hf}{{\textstyle{1\over 2}}}
\newcommand{\tr}{{\rm tr}}

\newcommand{\half}{{\textstyle{1\over 2}}}

\newcommand{\Z}{{\bf Z}}

\newcommand{\ra}{\rightarrow}

\newcommand{\delbar}{\overline{\partial}}
\newcommand{\zbar}{{\overline{z}}}

\newcommand{\Qbar}{{\overline{Q}}}

\def\a{\alpha} 
\def\b{\beta} 
\newcommand{\del}{\partial}
\def\G{\Gamma}
\def\e{\epsilon}

\newcommand{\cQ}{{G}}
\begin{document}
\addtolength{\baselineskip}{.5mm}
\newcommand{\dslash}{D}
\newcommand{\GG}{G}
\newcommand{\BPS}{{\mbox{\small\sc BPS}}}
\newcommand{\bps}{_{{}_{\rm BPS}}}
\newcommand{\LL}{_{\! {}_{L}}}
\newcommand{\RR}{_{\! {}_{R}}}
\newcommand{\pslash}{{p \hspace{-5pt} \slash}}
\newcommand{\aA}{{\bf a}}
\newcommand{\Ppsi}{{\bf \psi}}
\newcommand{\LR}{_{{}_{\! L,R}}}
\newcommand{\dnul}{\d_{{}_0}}
\newcommand{\lL}{{\mbox{\small  \cal L}}}
\newcommand{\gG}{{\mbox{\small \cal F}}}
\newcommand{\sS}{{\lambda}}
\newcommand{\sSbar}{{\overline\lambda}}
\newcommand{\Bps}{{\mbox{bps}}}
\newcommand{\xX}{x}
\newcommand{\Lup}{^{\! {}^{L}}}
\newcommand{\NP}{{N_P}}
\newcommand{\R}{{\bf R}}
\newcommand{\ebar}{{\overline\varepsilon}}
\newcommand{\nn}{k}
\newcommand{\Wedge}{\!\hspace{-1pt} \wedge \hspace{-1pt}}
\renewcommand{\e}{{\varepsilon}}
\newcommand{\cPo}{{H}}
\newcommand{\cPb}{{H}}
\newcommand{\ww}{{\bf 1}}

\begin{flushright}
April 1996\\
{\sc hep-th}/9604055\\
{\sc cern-th}/96-93\\
{\sc pupt}-1603
\end{flushright}
\vspace{-.3cm}
\thispagestyle{empty}

\begin{center}
{\Large\sc{BPS Quantization of the Five-Brane}}\\[13mm]

{\sc Robbert Dijkgraaf}\\[2.5mm]
{\it Department of Mathematics}\\
{\it University of Amsterdam, 1018 TV Amsterdam}\\[7mm]
{\sc Erik Verlinde}\\[2.5mm]
{\it TH-Division, CERN, CH-1211 Geneva 23}\\[.1mm]
and\\[.1mm]
{\it Institute for Theoretical Physics}\\
{\it University of Utrecht, 3508 TA Utrecht}\\[7mm]
and \\[3mm]
{\sc  Herman Verlinde}\\[2.5mm]
{\it Institute for Theoretical Physics}\\
{\it University of Amsterdam, 1018 XE Amsterdam} \\[.1mm]
and\\[.1mm]
{\it Joseph Henry Laboratories}\\
{\it Princeton University, Princeton, NJ 08544}\\[18mm]

{\sc Abstract}

\end{center}

\noindent 
We give a unified description of all BPS states of M-theory compactified
on $T^5$ in terms of the five-brane.  We compute the mass spectrum and
degeneracies and find that the $SO(5,5,\Z)$ U-duality symmetry
naturally arises as a T-duality by assuming that the world-volume
theory of the five-brane itself is described by a string theory. We
also consider the compactification on $S^1/\Z_2 \times T^4$, and give
a new explanation for its correspondence with heterotic string theory
by exhibiting its dual equivalence to M-theory on $K3\times S^1$.

\vfill

\newpage
\newsection{Introduction}

There is by now considerable evidence that the various dual relations
between different string theories can eventually be understood in
terms of an underlying ``M-theory'', whose low energy effective
action is given by eleven dimensional supergravity
\cite{hulltownsend,wittenm,duff,townsend,schwarz,horava,orbifolds,mem,hull11},
or its twelve-dimensional variant \cite{vafaF,kutasov}. It is not yet
known what M-theory looks like, but it is reasonable to expect that
just like string theory it has some formulation in terms of
fluctuating extended objects, being the membane or its dual, the
five-brane. The fundamental strings that we know in ten dimensions or
less are obtained by dimensional reductions from the membrane or the
five-brane.  From this point of view these branes must be considered
just as fundamental as strings \cite{townsend}. In particular, one
expects that upon quantization their spectrum will take the form of a
tower of states in an analogous fashion as for strings.

Unlike string theory, M-theory does not have a perturbative coupling 
constant, since there is no dilaton-field in 11-dimensional supergravity.  
The dual relationship between
membranes and five-branes must therefore be different from more
standard weak-strong coupling dualities, such as between strings and
five-branes in $d=10$. In particular, there is the logical possibility of a
double correspondence, in which the dual membrane may in fact also be
viewed as a particular limiting configuration of the five-brane
itself. Hence M-theory could in a sense be self-dual. 
This possibility is supported by the fact that, besides to the dual 
six-form $\tilde{C}_6$ of eleven-dimensional supergravity, the 
five-brane also couples directly to the three-form $C_3$ itself, via 
an interaction of the type
\be
\label{CT}
\int C_3\wedge T_3, 
\ee 
where $T_3$ is a self-dual three-form field strength that lives on the
world-volume. By allowing this field $T_3$ to have non-trivial fluxes
through the three-cycles on the world-brane, the five-brane can thus
in principle carry all membrane quantum numbers. These configurations
are therefore naturally interpreted as bound states between the two
types of branes.

One of the aims of the eleven-dimensional point of view is to shed
light on the mysterious U-duality symmetry of string theory
\cite{hulltownsend}.  The most convincing evidence for U-duality so
far has been obtained by considering the spectrum of BPS states. These 
studies necessarily involve D-branes that describe the states charged with
respect to the RR gauge fields \cite{polch}.  Indeed, there have been
convincing results in D-brane analysis, in particular in the form of
degeneracy formulas, that support the symmetry under certain U-duality
transformations \cite{vafa6d,sen}.  However, the formalism quickly
becomes rather involved, since in general one has to take into account
D-branes of various dimensions and also bound states of fundamental
strings and D-branes \cite{witten-b}.

As we just argued, the eleven-dimensional five-brane is a natural
candidate to give a more unified treatment of all BPS states in string
theory. In fact, one could hope that, in an appropriate covariant
quantization, U-duality becomes a manifest symmetry of the five-brane.
Unfortunately, there are various difficulties in extending the
covariant formalism from strings to higher-dimensional extended
objects \cite{bergshoeffetal}.  In this paper we will make a first
step in developing a formalism for describing the BPS configurations
of the five-brane for compactifications down to six dimensions, which
indeed exhibits the maximal symmetry.  An important ingredient in this
formalism is the idea that the relevant degrees of freedom on the
five-brane are formed by the ground states of a string theory living
on the world-volume itself. In fact, in compactifications on a 5-torus
the U-duality group $SO(5,5,\Z)$ can then be identified with the
T-duality group of this string. 

In six-dimensional compactifications only the ground states of this
string give rise to space-time BPS states. The string excitations
appear only after further compactification down to five and four
dimensions, where one can consider BPS representations that are
annihilated by 1/8 instead of 1/4 of the supercharges. The structure
of the resulting BPS spectrum has been described in our previous paper
\cite{letter}. In this paper we will restrict ourselves to the
six-dimensional theory.

\newsubsection{Outline of the paper}

In section 2 we will start with a detailed description of the BPS
spectrum of M-theory in six dimensions. First we derive the BPS mass
formula from the space-time supersymmetry algebra. Then, by comparing
with the known result of the BPS spectrum of type IIA strings, we
propose an explicit formula for all the degeneracies that is
manifestly invariant under the complete U-duality group $SO(5,5,\Z)$.
In section 3 we analyze the zero-mode structure of the five-brane and
show that the central charges correspond to particular fluxes through
homology cycles on the five-brane. We use this in section $4$ to
construct the complete space-time supersymmetry algebra, including the
central charges, as operators in the Hilbert space of the
five-brane. In section 5 we first argue that U-duality implies that
the world-brane theory must contain string degrees of freedom. We then
describe a possible light-cone formulation of this six-dimensional
string theory. The low-energy fields will correspond to the collective
modes of the five-brane.  Finally, in section 6 we set out to
calculate the BPS spectrum of the five-brane, first by considering the
linearized quantum fluctuations and then by including the winding
quantum numbers. We also discuss the relation with more conventional
D-brane counting. In section 7 we consider M-theory compactifications
on the orbifolds $S^1 \times K3$ and $T^4 \times S^1/\Z_2$. Here we
find a new derivation of the correspondence with heterotic string
theory and demonstrate the dual equivalence of these two
compactifications.

\pagebreak

\newsection{The BPS Mass Formula and U-duality}

Before we turn to our discussion of the five-brane theory, let us first
give a description of the BPS states and the mass formula from the six 
dimensional space-time point of view.  The maximally extended 
six-dimensional $N=(4,4)$ supersymmetry algebra is given by 
\cite{salamsezgin}
\ba
\left\lbrace Q^a_\alpha,Q^b_\beta\right\rbrace\is \omega^{ab}
\gamma^\mu \pslash_{\alpha\beta},\nonu \left\lbrace
Q^a_\alpha,\Qbar^b_{\overline{\beta}}\right\rbrace \is
\delta_{\a\overline\b} Z^{ab} ,
\ea 
where $a,b = 1,\ldots,4$ are $SO(5)$ spinor indices and $\omega^{ab}$
is an anti-symmetric matrix, that will be used to raise and lower
indices. (We refer to the Appendix for our conventions on spinors and
gamma-matrices.) The algebra contains 16 central charges that are
combined in the $4\times 4$ matrix $Z^{ab}$ and transform as a
bi-spinor under the $SO(5)\!\times\!SO(5)$ R-symmetry. It further
satisfies the reality condition $Z^* =\omega Z\omega^T$. We now wish
to obtain a convenient expression of the masses of BPS states in terms
of the matrix $Z^{ab}$.

In general, BPS states form short multiplets of the supersymmetry
algebra which are annihilated by a subset of the supersymmetry
generators. We will see that in six dimensions the generic BPS state
is annihilated by 1/4 of the 32 supercharges. For a given multiplet
the condition can be written as
\be
\label{BPS}
(\e_a Q^a + \ebar_b \Qbar^b)|\mbox{\small\sc
BPS}\rangle = 0 .
\ee 
Since this condition holds with fixed $\e,\ebar$ for all states in the
BPS multiplet, we can take the commutator with the supercharges, and
derive the following conditions on the supersymmetry parameters 
\ba
\pslash\e^a + Z^{ab}\overline{\e}_b \is 0,\nonu 
\pslash\overline{\e}_a + Z^\dagger_{ab}\e^b \is 0 .
\ea 
Combining these equations with the mass shell condition $p^2 +
m^2\bps=0$, one deduces that $m\bps^2$ coincides with the highest
eigenvalue of the hermitean matrices $ZZ^\dagger$ and $Z^\dagger\!Z$,
with $\e$ and $\ebar$ being the corresponding eigenvectors,
\ba
\label{ev}
(Z Z^\dagger)^a{}_b \e^b \is m^2\bps\e^a, \nonu (Z^\dagger
\!Z )_a{}^b\, \overline{\e}_b \is m^2\bps \overline{\e}_a.
\ea 
This determines the BPS masses completely in terms of the central
$Z^{ab}$.

The number of states within a BPS supermultiplet are determined by the
number of eigenvectors with the highest eigenvalue $m^2\bps$. In fact,
one can show that for a given (non-zero) value of $m\bps$ there are
always two independent eigenvectors $\e$ and $\ebar$. This can be
seen, for example, by decomposing the matrices $ZZ^\dagger$ and
$Z^\dagger\!Z$ in terms of hermitean gamma-matrices $\Gamma_m$,
$m=1,\ldots,5$ as
\ba
\label{gammaex}
Z Z^\dagger \is m_0^2 \, \ww + 2 K\LL^m \Gamma_m, \nonu
Z^\dagger\!Z\, \is m_0^2\, \ww + 2 K\RR^m \Gamma_m.
\ea
These relations, which define $m_0^2$ and the 5-vectors $K\LL$ and
$K\RR$, directly follow from the reality condition on $Z^{ab}$.  The
eigenvalue equation for the spinors $\e$ and $\ebar$ are now replaced
by the Dirac-like conditions
\ba
\label{diraclike}
(K\LL\! \cdot\G - |K\LL|)\e \is 0,\nonu 
(K\RR\! \cdot\G - |K\RR|)\ebar\is 0,
\ea 
which have indeed (generically) two independent solutions.  Thus the
BPS condition (\ref{BPS}) can be imposed for 8 of the 32 supercharges.
Consequently, a BPS supermultiplet contains $(16)^3$ states: this is
in between the size of a massless multiplet, which has $(16)^2$
states, and that of a generic massive supermultiplet with $(16)^4$
states.

The ten components $(K\LL,K\RR)$ can be understood as follows. We have
seen that the 16 components of the central charge can naturally be
combined into a spinor $Z$ of $SO(5,5)$. Out of two such spinors we
can construct in the usual way a ten-dimensional vector with
components $K^m\LL ={\textstyle{1\over 8}}\tr(\Gamma^m Z Z^\dagger)$
and $K^m\RR = {\textstyle{1\over 8}}\tr(\Gamma^m Z^\dagger Z)$ as
introduced in (\ref{gammaex}). In fact, these quantities form a {\it
null} vector $(K\LL, K\RR)$ of $SO(5,5)$, since one easily verifies
that $|K\LL|=|K\RR|$.  We can express the BPS masses in terms of this
vector, by combining the results (\ref{ev}) and (\ref{gammaex}). We
find
\be 
\label{mbps}
m^2\bps = m_0^2 + 2 |K\LR|, 
\ee 
where $m_0^2={\textstyle{1\over 4}}\tr( Z Z^\dagger)$.  The above BPS
mass formula is invariant under an $SO(5)\times SO(5)$ symmetry, which
acts on $Z_{ab}$ on the left and on the right respectively.

\newsubsection{U-duality invariant multiplicities of BPS states}

Charge quantization implies that the central charge $Z^{ab}$ is a
linear combination of integral charges. The expression of $Z^{ab}$ in
terms of the integers depends on the expectation values of the 25
scalar fields of the 6-dimensional $N=(4,4)$ supergravity theory.
{}From the point of view of eleven dimensions these scalars represent
the metric $G_{mn}$ and three-form $C_{mnk}$ on the internal manifold
$T^5$, and parametrize the coset manifold ${\cal M} =
SO(5,5)/SO(5)\!\times\!  SO(5)$. A convenient way to parametrize this
coset is to replace the 3-form $C_3$ on $T^5$ by its Hodge-dual $B =
*C_3$. The parametrization of ${\cal M}$ in terms of $G_{mn}$ and
$B_{mn}$ is then familiar from toroidal compactifications in string
theory.  Infinitesimal variations of $G$ and $B$ are represented via
the action of the spinor representation of $SO(5,5)$ on
$Z^{ab}$. Concretely,
\be
\label{vari}
\delta Z^{ab} = (\delta G^{mn} +\delta B^{mn}) (\Gamma_m Z \Gamma_n)^{ab},
\ee
where $\Gamma_m$ are hermitean gamma-matrices of $SO(5)$.  The
U-duality group is now defined as those $SO(5,5)$ rotations that map
the lattice of integral charges on to itself. Thus the U-duality group
can be identified with $SO(5,5,\Z)$.

The 16 charges contained in $Z_{ab}$ can be interpreted in various
ways depending on the starting point that one chooses. In this paper
we will be interested in the BPS states that come from the five-brane
in 11 dimensions. From the point of view of the five-brane it is
natural to break the $SO(5,5,\Z)$ to a $SL(5,\Z)$ subgroup, because
this represents the mapping-class group of the five-torus $T^5$.  The
16 charges split up in 5 Kaluza-Klein momenta $r^m$, 10 charges
$s_{mn}$ that couple to the gauge-fields $C_\mu^{mn}$ that come from
the 3-form, and one single charge $q$ that represents the winding
number of the five-brane. We can work out the BPS mass formula in
terms of these charges $q$, $r^m$ and $s_{mn}$. To simplify the
formula we consider the special case where the scalar fields
associated with three-form $C_{mnk}$ are put to zero, and the volume
of $T^5$ is put equal to one. Note that these restrictions are
consistent with the $SL(5)$ symmetry. For this situation the central
charge $Z_{ab}$ takes the form
\be 
Z_{ab}= q \ww_{ab} + r^m \Gamma_{m,ab} + s_{mn} \Gamma^{mn}_{ab}, 
\ee 
where the Dirac-matrices satisfy
$\{\Gamma_m,\Gamma_n\}=2G_{mn}$. (Note that in our notation
$\ww_{ab}=\omega_{ab}$.)  Inserting this expression in to the BPS mass
formula gives
\be 
m\bps^2 = m_0^2 + 2 \sqrt{G_{mn}K^mK^n + G^{mn}W_mW_n},
\ee 
with 
\ba 
m_0^2 \is q^2 + G_{mn} r^m r^n+ G^{mk} G^{nr} s_{mn} s_{kr}, \nonu 
K^m \is  q\,r^m+ \hf \epsilon^{mnklr}s_{nk}s_{lr}, 
\label{KM}\\[2mm] 
W_m \is  s_{mn} r^n. \nonumber
\ea 
Here we have written $K\LR^m = K^m \pm G^{mn} W_n$.

The BPS spectrum has the following interpretation in terms of the
toroidal compactification of the type IIA string. Because the string
coupling constant coincides with one of the metric-components, say
$G_{55}$, string perturbation theory breaks the $SO(5,5,\Z)$ U-duality
to a manifest $SO(4,4,\Z)$ T-duality. Accordingly, the 16 charges split
up in an $SO(4,4)$ vector of NS-charges, being the 4 momenta and 4
string winding numbers $n^i$ ($=r^i$), and $m_i$ ($=s_{i5}$), and an 8
dimensional spinor that combines the RR-charges $q$, $s_{ij}$ and $r$
($=r_5$) of the 0-branes, 2-branes and 4-branes. U-duality relates
RR-solitons to the perturbative string states, and can thus be used to
predict the multiplicities of the solitonic BPS states from the known
spectrum of string BPS states. This fact has been exploited by Sen and
Vafa \cite{sen,vafa6d} to give a non-trivial check on U-duality by
reproducing the expected multiplicities from the D-brane description
of the RR-solitons \cite{polch}.  The multiplicities of the
string BPS states, \ie with vanishing RR-charges, is given by $d(N)$
where $N = n^i m_i$ and
\be
\sum_N d(N) t^N = (16)^2 \prod_\nn \left({1+t^\nn\over 1-t^\nn}\right)^8.
\ee
This formula also describes the degeneracies of the RR-solitons where
$N=qr+ \hf s_{ij}\tilde{s}^{ij}$ represents the self-intersection
number of the D-branes. In fact, with the help of our analysis, 
it is not difficult to obtain a generalized
formula that satisfies all requirements: we find that the degeneracies
are given by the same numbers $d(N)$, but where $N$ is now given by
the greatest common divisor of the ten integers $ K^m$ and $W_m$,
\be
N= {\rm gcd}({K^m, W_n}).
\ee  
Indeed, one easily verifies that for the string BPS states all
integers $K^m$ and $W_m$ vanish except $W_5=n^im_i$, while for the
RR-solitons the only non-vanishing component is $K^5= qr+\hf
s_{ij}\tilde{s}^{ij}$.  Furthermore, the formula is clearly invariant
under the U-duality group $SO(5,5,Z)$. It can be shown that this is
the unique degeneracy formula with all these properties.\footnote{This 
follows from the fact that any two
primitive null vectors $v,w\in \Gamma^{5,5}$ can be rotated into each other
by a $SO(5,5,\Z)$ transformation. This is implied by the isomorphism
$v^\perp/\langle v\rangle \cong \Gamma^{4,4}$, {\it i.e.} the little
group of $v$ is always $SO(4,4,\Z)$. We thank G.\ Moore and C.\ Vafa
for discussions on this point.}

In the remainder of this paper we will present evidence that {\it all}
BPS states can be obtained from the five-brane. To this end we will
propose a concrete quantum description of the five-brane dynamics that
reproduces the complete BPS spectrum, including the above degeneracy
formula.

\newpage 

\newsection{Charges and Fluxes on the Five-Brane}

In this section we consider the compactification of the five-brane
coupled to eleven-dimensional supergravity to six dimensions on a
five-torus $T^5$.  {}From its description as a soliton it is known
that, after appropriate gauge-fixing, the five-brane is described by
an {\it effective} world-brane theory consisting of five scalars, an
anti-symmetric tensor with self-dual field strength $T=dU$ and 4
chiral fermions \cite{callanetal}.  These fields, which parametrize
the collective modes of the five-brane solution, form a tensor
multiplet of the chiral $N=(4,0)$ supersymmetry on the world-brane.

The five-brane couples directly to the six-form $\tilde{C}_6$, the
metric and the three-form $C_3$, and hence after dimensional reduction
it is charged with respect to all the corresponding 16 gauge fields
that we denote as $A, A^m$ and $A^{mn}$.  To make this concrete, we
consider a five-brane with topology of $T^5\times \R$ where $\R$
represents the world-brane time $\tau$.  First we consider the
coupling to $A^{mn}$, which is deduced from the term (\ref{CT}) by
taking $C_3 = A_{mn}\!\wedge\!  dX^m\!\wedge\!  dX^n$, where $X^m$ are
the embedding coordinates of the five-brane, and $A^{mn}$ is constant
along $T^5$.  Now we use that $dX^m\!\wedge\!  dX^n$ represents a
closed two-form, and hence defines a dual three-cycle $T^3_{mn}$.  In
this way we find that the charge $s_{mn}$ with respect to the gauge
fields $A^{mn}$ is given by the flux of the self-dual three-form field
strength $T=dU$
\be 
s_{mn} = \int_{T^3_{mn}} dU 
\ee 
through the 10 three-cycles $T^3_{mn}$ on the five-brane. A five-brane
for which these charges are non-zero, is actually a bound state of a
five-brane with a number of membranes: the quantum numbers $s_{mn}$
count the number of membranes that are wrapped around the $2$-cycle
$T^2_{mn}$ that is dual to $T^3_{mn}$. This is similar to Witten's
description of bound states of $(p,q)$-strings \cite{witten-b}.
Notice that in this case that the zero-mode of the canonical momentum
$\pi_U$ is also quantized, but these are not independent because the
self-duality condition implies that $\pi_U = dU$.

Our aim in this paper is to arrive at a U-duality invariant
description of the five-brane. We have shown that under the U-duality
group $SO(5,5,\Z)$ the charges $s_{mn}$ are part of an irreducible
16-dimensional spinor representation together with the Kaluza-Klein
momenta $r^m$ of the five-brane and its winding numer $q$ around the
five-torus. In fact, the spinor representation of $SO(5,5)$ is
naturally identified with the odd (co)homology of the five-torus
$T^5$. This observation motivates us to try write the other charges
$q$ and $r^m$ as fluxes of 5- and 1-form field strengths through the
5-cycle and 1-cycles on $T^5$.  Indeed, the winding number of the
five-brane around the internal $T^5$ can be written as 
\be
\label{qflux}
q=\int_{T^5} dV,
\ee 
where $dV$ is interpreted as the pull-back of the constant
volume element on the $T^5$-manifold.  We now would like to turn the
4-form potential $V$ into an independent field that is part of the
world-brane theory.\footnote{In fact, in a light-cone formalism for extended 
objects one naturally obtains a
residual gauge symmetry under volume preserving diffeomorphisms
\cite{hoppe}. For our five-brane these can be used to eliminate 
all dependence on the five compact embedding coordinates $X^m$ except
the volume-form $dV\!=\!dX^m\!\wedge\! dX^n \!\wedge\!dX^k \!\wedge\!
dX^l\!\wedge\!  dX^p\epsilon_{mnklp}$. Hence, the five spatial
components $*V^m$ are basically the embedding coordinates
$X^m$. Notice also that the gauge-transformation $V\ra V+d\Lambda$
corresponds to a volume preserving diffeomorphism on the
five-brane, since it leaves $dV$ invariant.}

As mentioned, the effective action on the world-brane contains as
bosonic fields, besides the tensor field $U$, five scalars. In the
following we will interpret four of these as being the transversal
coordinates in space-time. This leaves us with one additional 
scalar $Y$. In particular, since the world-brane is 6-dimensional, we
can dualize it and obtain a four-form $V$. We will now identify this
field with the field $V$ in (\ref{qflux}). Its five-form field
strength $W=dV$ is normalized such that the flux $q$ is integer, but
for the rest it can take any value including zero. Now, formally we can
go back to the description in terms of the dual scalar field $Y$ by
taking $W$ to be an independent five-form and introducing $Y$ as the
Lagrange multiplier that imposes the Bianchi identity $dW=0$. The fact
that $W$ has integral fluxes implies that $Y$ must be a periodic
field, \ie $Y \equiv Y+r$ with r integer.  Notice that on-shell $dY =
\delta S/\delta\dot{V}=\pi_V$, and so $dY$ is the canonical momentum
for $V$.  The 5 remaining charges $r^m$ can now be identified with the
integer winding numbers of $Y$ around the 5 one-cycles $T^1_m$: 
\be
\label{momentu}
r^m =\int_{T^1_m} dY.  
\ee 
It is straightforward (see also the previous footnote) to show that
these operators are the generators of translations along the internal
directions on the 5-torus, and so they indeed represent the
Kaluza-Klein momenta of the five-brane.  In this way all the 16
charges $(q,r^m,s_{mn})$ have been written as fluxes through the odd
homology cycles on the five-brane, and so, in view of our previous
remark, are naturally identified with a 16 component $SO(5,5,\Z)$
spinor.

To make this more manifest it is convenient to combine the fields $Y$,
$U$ and $V$, and their field strengths $dY, dU$ and $dV$ using $SO(5)$
gamma-matrices as
\ba
\label{DY}
Y_{ab} \is Y \ww_{ab} + \ast V_m \G^m_{ab} + U_{mn}\G^{mn}_{ab},
\nonu (\nabla Y)_{ab} \is \ast dV \ww_{ab} + (d Y)_m \G^m_{ab} +
(*dU)_{mn}\G^{mn}_{ab},
\ea 
where $\nabla_{ab}= \Gamma_{ab}^m\partial_m$.  The field $Y_{ab}$ is
not an unconstrained field, since it would describe too many degrees
of freedom. Namely, we still have to impose the condition that $dU$ is
self-dual and $dY$ is the dual of $dV$. This can be done in a rather
elegant way in a canonical formalism be imposing\footnote{This
generalizes the condition of self-duality to interacting theories,
since we do not need to assume that the field is described by a free
action.}  the constraint $\pi_Y^{ab}= \nabla Y^{ab}$, where
$\pi_Y^{ab}$ is the canonical conjugate momentum of $Y_{ab}$. This
constraint reduces the number of on-shell degrees of freedom from $8$
to $4$.

The advantage of the notation (\ref{DY}) is that it makes
the action of the U-duality group more manifest: the action of
$SO(5)\times SO(5)$ is from the left and right respectively, while the
other generators of $SO(5,5)$ act as in (\ref{vari}).  The results of
this section can now be summarized by the statement that the central
charge $Z_{ab}$ coincides with the zero-mode part of $(\nabla
Y)_{ab}$.

\newsection{Space-time Supersymmetry}

Our aim in this section is to investigate the BPS spectrum from the
view point of the world-brane theory.  At present we do not know a
consistent quantum theory for five-branes that is derived from a
covariant world-volume action. Fortunately, our only aim is to study
the quantum states of the five-brane that are part of the space-time
BPS spectrum, and, as we will see, for this we do not need to consider
the full five-brane dynamics. Furthermore, even without using the
details of the world-brane theory one can already say a lot about its
quantum properties just on the basis of symmetry considerations and
other general principles.  Our only assumption is that the five-brane
theory permits a light-cone gauge, so that there are $4$ transversal
coordinates $X^i$ in the 6 uncompactified dimension. In the following
sections this assumption will be justified by the fact that from this
starting point we are able to derive a Lorentz invariant BPS spectrum.

In the light-cone gauge the $SO(5,1)$ space-time Lorentz group is
broken to the $SO(4)$ subgroup of transversal rotations. On the
world-brane this group plays the role of an R-symmetry. We organize
the fields accordingly using $SO(4)$ representations with $\a,\dot\a$
indicating the two chirality spinors.  In addition our fields may
carry one or two spinor indices of the $SO(5)$ group of spatial
rotations on the five-brane. These are denoted by $a$, $b$, etc.  In
this notation we have the following fields on the five-brane 
\be
X^{\a\dot\b}, \psi^\a_b, \psi^{\dot\b}_a, Y_{ab}, 
\ee 
where $Y_{ab}$ is the field that we introduced in (\ref{DY}).  Notice
that each of these fields has four on-shell components.  These fields
represent the collective modes of the five-brane and their zero-modes
will be used to construct the space-time supersymmetry algebra. More
precisely, each of these fields has a canonical conjugate momentum
$\pi_X^{\a\dot\b}, \pi_\psi^{\a b}, \pi_\psi^{\dot\b a}$ and
$\pi_Y^{ab}$. Their zero-modes
\be 
p^{\a\dot\b}, S^\a_b, S^{\dot\b}_a, Z_{ab} 
\ee 
enter in the $N=(4,4)$ space-time supersymmetry algebra respectively
as the transversal momentum, part of the space-time supercharges
(namely those that are broken by the five-brane) and the central
charge.

The world-brane theory carries a chiral $N=(4,0)$ supersymmetry that
is generated by a set of supercharges $\cQ^\a_a$ and $\cQ^{\dot\a}_a$.
These supercharges satisfy in general the commutator algebra
\ba
\{\cQ^\a_a,\cQ^\b_b\} \is 
2 \epsilon^{\a\b}(\ww_{ab} \cPo + \G^m_{ab}(\cP_m + W_m)),\nonu
\{\cQ^{\dot\a}_a,\cQ^{\dot\b}_b\} \is 
2 \epsilon^{\dot\a\dot\b}(\ww_{ab} \cPo + \G^m_{ab}(\cP_m - W_m)).
\label{wbsusy}
\ea
Here $\cPo$ is the Hamiltonian on the five-brane, and $\cP_m$ are
world-brane momentum operators that generate translations in the five
spatial directions.  It will sometimes be convenient to combine them
into the matrix
\be
\cP_{ab} = \ww_{ab} \cPo + \G^m_{ab}\cP_m. 
\ee
The operators $\cP_m$ play the same role as $L_0-\overline{L}_0$ in
string theory, and as is clear from this analogy, will have to
annihilate the physical states in the spectrum of the five-brane \be
\label{levelmatch}
\cP_m|{\rm phys}\rangle =0.  
\ee 
The operator $W_m$ that appear in the supersymmetry algebra
(\ref{wbsusy}) represents a possible vector-like central charge.  In
terms of the fluxes of $dY$ and $T=dU$ it receives a contribution
given by the topological charge
\be
W_m = \int_{T^4_m} dY \wedge T.
\ee
In order to have a realization of space-time supersymmetry (without
space-time vector central charges)
we will also have to put $W_m$ to zero on physical states
\be
W_m|{\rm phys}\rangle = 0.
\ee
If we do not introduce extra degrees of freedom, this condition implies
the relation $K\LL=K\RR$. Finally, from the
light-cone condition $x^+ =p^+\tau$, we find that we have to impose the 
condition  
\be 
\label{masshel}
(\cPo -p^+p^-)|{\rm phys}\rangle =0 .
\ee 

We will now turn to a discussion of the space-time supersymmetry.  To
reduce the number of equations somewhat, we will concentrate first on
the ``left-movers'' $\cQ^\a_a$, and discuss the dotted
``right-moving'' components $\cQ^{\dot\a}_a$ afterwards.  The
world-brane supercharges represent the unbroken part of the full
$N=(4,4)$ space-time supersymmetry algebra. The other generators must
be identified with the zero-modes $S^\a_{a}$ and $S^{\dot\a}_{a}$ of
the conjugate momenta of the world-brane fermions, since these are the
Goldstone modes associated with the broken supersymmetry. Under the
world-brane supersymmetry these zero-modes transform into the
zero-modes of the bosonic fields,
\ba
\{\cQ^\a_a,S^\b_b\} \is \epsilon^{\a\b} Z_{ab}, \nonu
\{\cQ^\a_a,S^{\dot\b}_b\} \is \omega_{ab} \pslash^{\a\dot\b}.
\label{first}
\ea 
To get the complete set of relations one needs to use the world-brane
supersymmetry algebra (\ref{wbsusy}) together with the conditions
(\ref{masshel}) and (\ref{levelmatch}). This gives on physical states
\ba
\{\cQ^\a_a,\cQ^\b_b\} \is 2 p^+p^-\epsilon^{\a\b} \omega_{ab}, \nonu
\{S^\a_a,S^\b_b\} \is \epsilon^{\a\b}\omega_{ab} ,
\label{second}
\ea 
and similarly for the dotted components. The space-time supersymmetry
generators are therefore
\be
Q^a_\a = (\mbox{\small $\sqrt{2 p^+}$} S^{\dot\a}_a, 
\cQ^\a_a/\mbox{\small $\sqrt{p^+}$}), 
\ee
where on the left-hand side $\a$ denotes a chiral four-component
$SO(5,1)$ space-time spinor index.

We can now discuss the space-time BPS states from the world-brane
point of view.  The value of the BPS mass is determined by the central
charge $Z_{ab}$, and so we know that for a BPS state we should find
that
\be 
\cPo|\BPS\rangle = \hf (p_i^2+m\bps^2)|\BPS\rangle 
\ee 
where $m\bps^2$
is given in (\ref{mbps}). We will now prove that this is in fact the lowest
eigenvalue of the Hamiltonian $\cPo$ in the sector with a given central
charge $Z_{ab}$. For this purpose, let us introduce new operators
$\hat\cQ^\a_a$ and $\hat{\cP}_{ab}$ which are defined as the non-zero
mode contributions of the respective operators. They satisfy the relations 
\ba
\cQ^\a_a \is Z^\dagger_a{}^bS^\a_b +\pslash^{\a\dot\b} S_{\dot\b,a}+
\hat\cQ^\a_a , \nonu 
\cP_{ab} \is \hf p_i^2 \delta_{ab} + \hf (Z^\dagger Z)_{ab}
+\hat{\cP}_{ab}. 
\ea 
Using the world-brane and space-time supersymmetry
algebra we derive that the anti-commutator of
two of these operators is 
\be 
(\hat\cQ^\dagger \hat\cQ)_{ab}\equiv
\half
\epsilon_{\a\b}\{\hat\cQ^\a_a,\hat\cQ^\b_b\}= 2\cP_{ab} -  p_i^2\delta_{ab}-
 (Z^\dagger Z)_{ab}. 
\ee 
Next, by taking the trace with the non-negative definite matrix
$\rho=\hf(1+\hat{K}\cdot\Gamma)$, where $\hat{K}$ is a unit vector in
the direction of $K=K\LL=K\RR$, one deduces that
\be
\cPo+\hat{K}\cdot \cP =  \half p_i^2 +  \half 
m\bps^2 + \half \tr(\rho\, 
\hat\cQ^\dagger \hat\cQ). 
\label{bound}
\ee 
The last operator on the right-hand-side is clearly positive
definite. Furthermore, since BPS states are physical, they have to be
annihilated by the translation generators $\cP^m$. Combining these two
facts gives the statement we wanted to prove. It also tells us that
BPS states are annihilated by half of the operators $\hat\cQ_a$
\be 
\e^a\hat\cQ_a|\BPS\rangle=0. 
\ee 
In this sense, space-time BPS states are also BPS states from
the world-brane point of view.

\newsection{Strings on the Five-Brane}

The U-duality group acts on the shape and size of the internal 5
torus, and in particular contains transformations that map large to
small volumes.  If we require that the five-brane theory is invariant
under such transformations, it is clearly necessary to include short
distance degrees of freedom. Furthermore, these extra degrees of
freedom need to behave the same for very small box sizes as momentum
modes do for large box sizes.  This suggests that we can possibly
restore the full U-duality invariance by replacing the world-brane
theory by a string theory. Another independent indication that the
five-brane world-volume theory may in fact contain string-like
excitations is the possible occurrence of vector-like central charges 
in the supersymmetry algebra, since only one-dimensional extended 
objects can carry such charges.

One easily sees that BPS states necessarily correspond to the string
ground states. These states are annihilated by 1/4 of the {\it space-time}
supercharges, which implies, as we have seen in the previous section,
that they are annihilated by 1/2 of {\it five-brane} supercharges. If we
repeat this procedure once more, we conclude that in terms of the
string, BPS states are annihilated by all of the {\it string}
supercharges. Although we only use the string ground states in this paper
(see however \cite{letter}) we will now make some remarks concerning
the formulation of this six-dimensional string theory. 
 
We are looking for a string model whose ground states represent the
massless tensor multiplet describing the collective modes of the
five-brane.  Specifically, we expect ground states of the form
$|\alpha\dot\beta\rangle$ that describe the four scalars
$X^{\alpha\dot\beta}=\sigma_i^{\alpha\dot\beta} X^i$, states $|\alpha
b\rangle$ and $|a\dot{\beta}\rangle$ that describe the 4 world-brane
fermions $\psi^\alpha$ and $\psi^{\dot\alpha}$, and finally we need
states $|ab\rangle$ that correspond to the fifth scalar $Y$ and the 3
helicity states of the tensor field $U$. Here the indices $a,b=1,2$
now label chiral $SO(4)$ spinors.\footnote{Though the string theory
we want needs to contain the tensor field $U$ in its spectrum, note that
it should not carry any charge with respect to it, since this would
violate charge conservation on the five-brane. Rather, the $U$-field should
couple via its field strength $T$, like an RR-vertex operator. 
The string described in
this section should therefore be distinguished from the self-dual
string considered in \cite{sdstring}. We thank M. Becker, J. Polchinski,
and A. Strominger for discussions on this point.}

This structure arises naturally in the following model.  We will
assume that the world-sheet theory of this string theory can be
formulated in a light-cone gauge, and so one expects to have $4$
transversal bosonic coordinate fields $\xX^{a\dot a}(z,\zbar)$
together with fermionic partners $\sS^a_\a(z),$
$\sSbar^a_{\dot a}(\zbar)$.  The world-sheet theory has 4 left-moving
and 4 right-moving supercharges $\gG^{\dot{a}\a} =\oint
\del \xX^{\dot a b} \sS_b^\a$ and $\overline{\gG}^{\dot a \dot\a}=\oint
\delbar \xX^{\dot a b} \sSbar_b^{\dot\a}$ which correspond to the
unbroken part of the world-brane supersymmetry and satisfy
\be
\{ \gG^{\dot a\a} ,\gG^{\dot b\b}\} = 
2\epsilon^{\dot a \dot b}  
\epsilon^{\a \b} \, \lL_{0},
\ee
where $\lL_{0}$ is the left-moving world-sheet Hamiltonian.
The ground states must form a multiplet of the 
zero-mode algebra $\{\sS_a^\alpha,\sS_b^\beta\}=
\epsilon_{ab}\epsilon^{\alpha\beta}$. 
This gives 2 left-moving bosonic ground states $|\alpha\rangle$ and 2
fermionic states $|a\rangle$. By taking the tensor product with the
right-moving vacua one obtains in total $16$ ground states
\be
\Bigl(|\a, k\LL\rangle \oplus |a,k\LL\rangle\Bigr)\otimes
\Bigl(|\dot\b, k\RR\rangle \oplus |b,k\RR\rangle\Bigr),
\ee
just as we wanted.
Here we also took into account the momenta $(k\LL, k\RR)$, which form
a $\Gamma_{5,5}$ lattice, since we have assumed that the five-brane
has the topology of $T^5$. Since the light-cone coordinates
parametrize a cylinder $S^1\times \R$, the standard light-cone
formalism has to be slightly modified, as we have to take into account
that the string can also wind around this circle. Specifically, we
must impose the mass-shell condition
\be
\lL_0=  k^+\LL k^-\LL = \half k_0^2 - \half (k\LL^5)^2,
\ee
so that the level matching condition between the left-moving and
right-moving sectors of the string becomes
\be
\lL_{0} - \overline{\lL}_{0} = \half (k\LL^5)^2 - \half (k^5\RR)^2 
= k^5 w^5.
\ee
with $w^5$ the winding number around the $S^1$.
Level matching implies that for the ground states $|k\LL|=|k\RR|$.

The world-sheet supersymmetry transformation relates the fermion
zero-modes and the transversal momentum $k\LL$ of the string
\be
\{ \gG^{\dot a\a} ,\sS^{b\b}\} = k\LL^{a\dot b} \epsilon^{\a\b}.
\ee
Hence one can construct a (4,0) supersymmetry on the 5+1-dimensional target
space of the string, \ie on the world-volume of the five-brane, as follows 
\be
\hat\cQ^{a\a} = ({\mbox{\small $\sqrt {2 k}{}\LL^+$}}  \sS^{a\a} , 
\gG^{\dot a\a}/{\mbox{\small$\sqrt {k}{}\LL^+$}}  ),
\ee
where on the left-hand side $a$ denotes a chiral SO(5,1) spinor index. 
These generate the algebra
\be
\{ \hat\cQ^{a\a} ,\hat\cQ^{b\b}\} =  2 \epsilon^{\a\b} 
( k^0 \ww^{ab} + k\LL^m \G_m^{a b})
\ee
and similar for the right-moving generators.

We now propose that the world-brane dynamics of a single five-brane is
described by the second quantization of this string theory.  The most
important consequence for our study of the BPS states is that the
world-brane supersymmetry algebra gets modified.  Namely, as we have
just shown, the anti-commutator of the left-moving supercharges
produces the left-moving momentum $k\LL$, while the right-movers give
$k\RR$.  Hence the string states form representations of the $N=(4,0)$
supersymmetry algebra\footnote{As explained in section 4, the
appropriate second-quantized generators $\cQ^{a\a}$ also contain a
zero-mode contribution in addition to the part $\hat\cQ^{a\a}$ which
is expressed in terms of the string creation and annihilation
operators.  We have silently added the zero-mode contribution here.}
\ba
\label{left}
\{\cQ^{a\a} ,\cQ^{b\b}\} \is 2
\epsilon^{\a\b}( \cPb \ww^{ab} + \cP\LL^m\Gamma_m^{ab}),
\nonu
\{ \cQ^{a\dot\a} ,\cQ^{b\dot\b}\} \is
2 \epsilon^{\dot\a\dot\b}( \cPb \ww^{ab} + \cP\RR^m\Gamma_m^{ab}).
\ea
The operators $\cPb$, $\cP\LL^m$ and $\cP\RR^m$ act on multi-string
states that form the Hilbert space of the five-brane.  The five-brane
Hamiltonian $\cPb$ measures the energy of the collection of strings,
while $\hf(\cP\LL^m +\cP\RR^m)$ measure the total momentum. But we see
that the algebra also naturally contains a vector central charge
$W^M = \hf(\cP\LL^{m}\! -\!
\cP\RR^{m})$, which measures the sum of the string winding numbers
around the 5 independent one-cycles on the world-brane.

We are interested in BPS states, which are annihilated by eight of the 
world-brane supersymmetry generators. Such states are obtained by
combining individual string states $|\Bps\rangle$ that satisfy
\be
\varepsilon_{a\a} \hat\cQ^{a\a} |\Bps\rangle = 0,
\ee
where $\e$ is constrained by
\be
k\LL^{ab} \varepsilon_{b\a} = 0,
\ee
and similarly for the opposite chirality.
In terms of the world-sheet generators this condition reads
\be
\gG^{a\a} |\Bps\rangle = k\LL^{ab}\sS_{b\a}|\Bps\rangle.
\ee
This condition tells us that the string must be in the
left-moving ground state. The full BPS condition thus implies that
the individual strings must be in their ground state. 

\newsection{U-duality Invariant BPS Spectrum of the Five-Brane}

We will now discuss the BPS quantization of the five-brane.  A natural
starting point is the low-energy effective field theory on the
world-volume, which is appropriate for describing the world-brane
dynamics at large volume. Our hypothesis is that the effective theory
in this regime takes the most simple form consistent with all the
symmetries of the five-brane. As we have argued, this minimal
requirement is fulfilled by the field theory consisting of the single
tensor multiplet described in sections 3 and 4. These fields describe
the ground states of the strings, except that at small volumes we have
to take into account the winding configurations of the underlying
string theory. We will further make the assumption that for the
purpose of constructing the BPS spectrum it is allowed to treat the
five-brane fluctuations in a linear approximation, that is, we will
use free field theory.

Since we will work towards a Hamiltonian formalism, we will from the
beginning distinguish the time-coordinate $\tau$ from the five
space-like coordinates $\sigma^m$. We can further pick a Coulomb gauge
for the two-form field $U$ by demanding $U_{m0}=0$ and 
$\partial^m U_{mn} = 0$. It is useful to introduce the matrix-valued 
derivative 
\be
\dslash_{ab} = \ww_{ab} \partial_0 + \G^m_{ab}\partial_m 
\ee 
and the world-volume bi-spinor field $Y^{ab}$ introduced in equation
(\ref{DY}). Its equation of motion then takes the form
\be
\label{eom}
D_a{}^b({D}Y)_{bc} = D_{a}{}^b(DY)^\dagger_{bc} = 0 ,
\ee 
where $D$ acts on $Y$ via matrix multiplication.  These two equations
reduce the number of independent on-shell components of $Y_{ab}$ to
four.  The free field equations of motion of the other fields are in
this notation 
\ba 
D^{ab}(DX)^{\a\dot\b}_{ab} \is 0, \nonu 
(D\psi)^\a_a \is 0 .
\ea 
The free field Hamiltonian and momentum operators on the world-brane
take the quadratic form
\be
\cP^{ab} = \int_{T^5} \!\! d^5\!\sigma\, \half \left[
(DY)^{ac}(DY)^\dagger_c{}^b+ (\dslash X_i)^{ac}
(\dslash X^i)_c{}^b
+ \psi^a_\a (\dslash\psi)^{b\a} + \psi{}^a_{\dot\a} (\dslash\psi)^{b\dot\a} 
\right]
\label{Pab}
\ee

To describe the quantum fluctuations of the five-brane, we expand the
various fields in plane waves that propagate on the
world-volume. Since we assume that the world-volume has the topology
of $T^5\times \R$, we can label these waves with 5 integral momenta $k_m$.
For example, the expansion of $Y_{ab}$ is 
\be
\label{exp}
(DY)_{ab} = Z_{ab} + \sum_k (\aA^\dagger_{I}(k) u^I_{ab}(k) e^{ik\cdot
\sigma} + \aA_{I}(k) \overline{u}_{ab}^I(k) e^{-ik\cdot \sigma}), 
\ee 
where
$I$ runs from 1 to 4, and $k$ runs over the five-dimensional momentum
lattice of $T^5$.  Together with the four other bosonic fields
$X^{\a\dot\b}$ and fermionic fields $\Ppsi_a^\a$ and
$\Ppsi_a^{\dot\a}$, we obtain creation and annihilation modes
$\aA_I(k)$, ${\Ppsi}_A(k)$, where the indices $I$ and $A$ now
both run from 1 to 8. 

We could now impose the BPS conditions in the light-cone formalism 
that we have been using as described in section 4. However, as we 
have seen, the effective field theory approach is only
sufficient (and consistent) in the special case that the central charge
satisfies the condition $W^m=\hf(K\LL^m \! - K\RR^m)=0$. 
This condition, which breaks U-duality,  can be understood as
the absence of vector central charges on the world-brane. In terms
of the string theory this corresponds to considering the sector with
zero total winding number. Instead of working out this case in detail,
we will immediately give a manifestly U-duality invariant derivation of
the BPS spectrum. The field theory limit can simply be obtained
afterwards by simply putting $W^m$ to zero.
 
Indeed, it is clear from the discussion in section 5 that a field
theory description is incomplete and that at small 
volumes of $T^5$ extra degrees of freedom must be included in the effective 
description. In particular, the presence of string winding states 
will imply that the modes of the quantum fields can carry 
besides the momentum $k$ also a winding number $w$. 
To put this idea into effect, we note that the equation of motion 
(\ref{eom}) of the field $DY$ in momentum space can indeed be 
generalized to an $SO(5,5)$ invariant equation by introducing 
a left- and a right-momentum vector, as follows
\ba
(|k| \ww - k\LL \! \cdot \Gamma)^{ab}\, 
u^I_{bc}(k\LL,k\RR)  \, \is 0, \nonu
(|k| \ww - k\RR \! \cdot \Gamma)^{ab}  
u^\dagger{}^I_{bc}(k\LL,k\RR) \is 0.
\ea
Here $u^I_{ab}(k\LL,k\RR)$ denotes the generalization of the bi-spinors 
$u^I_{ab}(k)$ in the mode expansion (\ref{exp}) of $DY_{ab}$. 
Via the action of $SO(5,5)$ on the spinor indices of $DY$, we deduce
that the pair of momenta $k\LL$ and $k\RR$ combine into an $SO(5,5)$ 
null-vector, with $k\LL^2 - k\RR^2 = 0$. In a similar way, we can
argue that all other fields must also depend on two momenta 
instead of one.

We are thus led to consider a Fock space with stringy oscillators 
$\aA^I(k\LL,k\RR),\Ppsi^A(k\LL,k\RR)$ where $(k\LL, k\RR)\in 
\G^{5,5}$ and $|k\LL|=|k\RR|= |k|$. The generalized number 
operator is 
\be
N_{k_L,k_R}= \aA^\dagger_I(k\LL,k\RR) \aA^I(k\LL,k\RR) +
\Ppsi_A^\dagger(k\LL,k\RR) \Ppsi^A(k\LL,k\RR).
\ee
We can now write the Hamiltonian and momentum operators on the 
five-brane (see equation (\ref{Pab}))
in terms of the contributions of the zero-modes and these 
particular string modes as
\ba
\label{hamilton}
\cPo \is \half m_0^2 + \half p_i^2 + \sum_{k_L,k_R} |k|N_{k_L,k_R}, \nonu 
\cP\LL^m\is K\LL^m+\sum_{k_L,k_R} k\LL \,N_{k_L,k_R}.
\label{momentum} 
\ea
where $m_0^2 = {1\over 4} \tr Z^\dagger Z$ and 
$K\LL^m = {1\over 8}{\rm tr}(\Gamma^m Z^\dagger\! Z)$. 
Similarly, we define $\cP\RR^m$ with $K\RR^m = {1\over 8}{\rm tr}
(\Gamma^m Z Z^\dagger)$. 

Note that we now have independent left-moving and right-moving
momentum operators $\cP\LL$ and $\cP\RR$ on the five-brane. In 
order to be able to realize the space-time supersymmetry algebra, 
we have to impose on physical states the conditions that both 
momentum operators vanish 
\be
\cP\LL^m|{\rm phys}\rangle = \cP\RR^m |{\rm phys}\rangle =0.
\label{level}
\ee
These equations tell us that the sum of the individual
left-moving or right-moving string momenta $k\LL,k\RR$ have to cancel the 
contribution $K\LL,K\RR$ of the zero-mode fluxes on the five-brane.
We see in particular that, in the case that $K\LL\neq K\RR$, this
necessitates the inclusion of string winding modes. 

We will now impose the BPS condition. Hereto we can use the result
demonstrated at the end of section 4, that BPS states saturate the
lower bound (\ref{bound}) for the ``light-cone Hamiltonian'' $\cPo +
\hat K \!\cdot\! \cP\LR$ for a given value of the central charge $Z_{ab}$.
In more detail we proceed as follows.  The (mass)${}^2$ of these
five-brane states is measured by the operator
\be
m^2= m_0^2 + \sum_{k_L,k_R} 2 |k| N_{k_L,k_R}.
\ee
As we have explained in detail in section 2, the BPS condition implies
that $m^2$ must be equal to $m^2\bps= m_0^2 + 2|K\LR| $ 
which is the highest eigenvalue of $Z^\dagger Z$ or 
$ZZ^\dagger$. Using the physical state condition $\cP\LL=0$, 
the BPS mass formula may be rewritten as
\be
\half m^2+ \hat{K}\LL \!\cdot\! P\LL =  \half m_0^2+|K\LL|,
\ee
where we introduced the unit vector $\hat{K}\LL$ in the direction of
$K\LL$.  Inserting the mode-expansions of $\cP\LL$ and $m^2$ shows
that the zero-mode part of the left-hand side is already equal to the
right-hand-side.  The remaining oscillator part must therefore add up
to zero
\be
\sum_{k_L,k_R} (|k| + \hat{K}\LL\!\cdot\! k\LL) N_{k_L,k_R} =0.
\ee
The left-hand side is a non-negative expression and can vanish only if
the excitations have momentum $k\LL$ directed in 
the opposite direction of $K\LL$. Similarly, one obtains an analogous
result for the right-moving momenta $k\RR$. So, we conclude that the 
BPS spectrum for given central charge vector $K=(K\LL,K\RR)\in\G^{5,5}$
is obtained by acting with only those oscillators
$\aA_k$ and $\Ppsi_k$ for which the momentum vector $k=(k\LL,k\RR)$ 
points in the direction opposite to $K$. 

To count the number of BPS states for given vector $K\in \G^{5,5}$, 
let $[K]$ be the largest positive 
integer so that $K/[K]$ is still an integral vector. 
In other words, $K$ is $[K]$ times a primitive vector, and the $[K]$ can be 
defined as the greatest common divisor of the ten integers $K\LL^m,K\RR^m$. 
With this notation, the allowed momenta of the oscillators must be of the 
form $k_n= - n K/[K]$, for some positive integer $n$. Let $N_n$ denote the 
occupation number of these modes. The level matching 
conditions (\ref{level}) together with (\ref{momentum}) now reduce to
the simple combinatorial relation 
\be
\sum_n n N_n = [K].
\ee
Since there are 8 bosonic and 8 fermionic modes that contribute at each 
oscillator level, we obtain the result that we announced at the end of
section 2. The number of BPS states is given by $d([K])$ with
\be
\sum d(N) t^N = (16)^2 \prod_n \left({1+t^n\over 1-t^n}\right)^8.
\ee   

We would like to point out that the above result has a rather striking 
interpretation. We see that in the BPS limit the excitations of the 
five-brane are constrained to lie in a single space-like direction,
which is determined by the value of the central charge. So, effectively
the six-dimensional world-volume reduces to a world-sheet and the 
BPS five-brane behaves like a string, in fact a chiral type II string.
The U-duality group $SO(5,5,\Z)$ acts on the momentum vector $K$ and
so permutes the various string-like excitations of the five-brane.

If we want to relate this eleven-dimensional point of view to the
ten-dimensional type II string, we have to single out a particular direction
on the world-brane. As we discussed before, this breaks the U-duality group
to the little group $SO(4,4,\Z)$ and corresponds to distinguishing 
NS-NS and R-R type charges. The string-like excitations in this
fixed direction then correspond directly to the BPS states 
of the fundamental type II string.

\newsubsection{Comparison with D-branes}

It might be illustrative to compare the counting of BPS states in this
paper with the more conventional counting using D-branes \cite{polch}.
It is especially instructive to see how the string degrees of freedom
of the five-brane, which from our point of view were a crucial
ingredient in a complete description of the BPS spectrum, manifest
themselves in the world-volume theories of the D-branes. In particular
we would like to see to which extent string-like excitations occur for
the Dirichlet four-brane in type IIA superstring theory, which can be
considered as a simultaneous dimensional reduction to ten dimensions
of the five-brane in M-theory.

So let us reconsider the compactification of the type IIA superstring
on a four-torus. As we mentioned in section 2, the 16 quantum numbers
now decompose in the 8 NS-NS and 8 R-R charges
\ba
(p,w) &\in & H^{odd}(T^4),\nonu
(q_0,q_2,q_4) & \in & H^{even}(T^4)
\ea
Here $p_i,w^i$ are the momenta and winding number of the fundamental
NS string, and $q_p$ $(p=0,2,4$) is the Dirichlet $p$-brane charge. In
terms of these charges the 10 components of the vector $(K^m,W_m)$ of
equation (\ref{KM}) are given by
\ba
K^5 \is q_4 q_0 + \half q_2\wedge q_2 \nonu
K^i \is q_4 p^i + (q_2 \wedge w)^i \nonu
W_5  \is p^i w_i \nonu
W_i \is q_0 w_i + (q_2 \cdot p)_i
\ea
Our analysis and U-duality predict that the degeneracy is a function
of the g.c.d.\ of these 10 integers.

Let us now look at configurations in which we can make a meaningful
comparison with standard D-brane computations to count bound state
degeneracies. We will be particularly interested in configurations
with a single 4-brane, no 2-brane (for simplicity) and an arbitrary
number of 0-branes, {\it i.e.}\ we put $q_4=1$ and $q_2=0$. In that
case the degeneracy is given by $d(N)$ with
\be
N = gcd(q_0,p^i,p\cdot w,q_0w^i)=gcd(q_0,p^i).
\ee

Can we make a macroscopic BPS string-state on the four-brane which has
a non-zero winding number $W^i$? Such an object will manifest itself,
in the large volume, as a long string-like object. In our philosophy
such a state is a coherent sum of BPS five-brane strings with paralel
momenta and winding numbers.

If we choose for simplicity the total momentum $K^i$ to be zero, we
see from the above expressions that it is indeed quite easy to make
such an object. For $q_0=1$ and $p^i=0$ the winding number $W^i$
simply equals the winding number $w^i$ of the fundamental NS string.
Furthermore, the degeneracy of such a state is given by the usual
number of string ground states $d(0)=2^8$. So in this case, the string
BPS state is nothing but the fundamental closed string bound with a
zero-brane to the four-brane.

In fact, this point of view can be easily generalized to a
configuration with an arbitrary number $q_0$ of zero-branes, where the
degeneracy is predicted to be $d(q_0)$. There is a simple explanation
of this counting using the so-called ``necklace'' model. In this case
we have $q_0$ zero-brane ``beads'' that are stringed together with a
NS string and bound to the four-brane. These zero-branes will cut the
string in $q_0$ pieces, each of which has the usual $8+8$ ground
states. However, the zero-branes can cluster together and form bound
states. Therefore, to compute the total number of BPS states we have
to sum over partitions giving the usual degeneracy formula
\be
\sum d(N) t^N = 2^8 \prod_n \left({1+t^n \over 1 - t^n}\right)^8
\ee

\newsection{Heterotic String Theory from the Five-Brane.}

The above results can be generalized in a straightforward fashion to
other internal manifolds than $T^5$. As an explicit example, we will
now briefly discuss the cases of $S^1 \times K3$ and $T^4 \times
S^1/\Z_2$.  Previous studies of M-theory compactified on these
manifolds \cite{horava,orbifolds,duffetal} have shown that in both
cases the theory becomes equivalent to the heterotic string
compactified on $T^4$ (provided that the $\Z_2$ action in the latter
case is defined appropriately). Here we confirm this by means of an
explicit study of the five-brane BPS spectrum. In particular we will
find that the two different compactifications are T-dual from the
point of view of the world-volume string theory.

It will be useful in the following to think about $K3$ as (a
resolution of) the orbifold $T^4/\Z_2$, so that both types of
compactification manifolds are obtained as $\Z_2$-orbifolds of $T^5$,
with the two $\Z_2$'s acting on $T^4$ and $S^1$ respectively.  These
two transformations are represented on the fluxes $Z^{ab}$ of
$DY^{ab}$ as follows
\be 
Z \ra \Gamma_5 Z \Gamma_5 
\ee 
and 
\be 
Z \ra - \Gamma_5 Z\Gamma_5 
\ee 
respectively.  To be able to construct the appropriate orbifolds we
will first have to extend these $\Z_2$-actions to a transformation on
the complete fields, and furthermore, check that the resulting
transformations are indeed symmetries of the five-brane theory.
Notice that under both $\Z_2$ actions the quantities $K^i$ and $W_i$
(with $i=1,\ldots,4$) are odd, while $K^5$ and $W_5$ are invariant.
This suggest that the $\Z_2$ action must be accompanied by a
reflection of the world-volume of the five-brane. More precisely, we
find that the unique $\Z_2$-symmetry of the world-volume action that
incorporates the above transformation on the fluxes is given by
the orientifold transformation
\ba 
DY(\sigma) & \ra & \pm \Gamma_5 \tilde{D}Y(\tilde{\sigma}) \Gamma_5, \nonu 
DX(\sigma) & \ra & DX(\tilde{\sigma}), \\[2mm] 
\psi(\sigma) & \ra & \Gamma_5 \psi(\tilde{\sigma}),\nonumber 
\ea 
where $\tilde{\sigma}^i = -\sigma^i$ for
$i=1,\ldots,4$, and $\tilde{\sigma}^5 =\sigma^5$. Without this coordinate
reflection this transformation would {\it not} be a symmetry. For
example, it would not leave the definition of the translation
operators $\cP_m$ invariant.

So for both types of orbifolds, the world-volume coordinates $\sigma$
lie on the same orbifold $S^1 \times T^4/\Z_2$, which we may also think
of as $S^1 \times K3$.  The two theories differ, however, via their boundary
condition on the field $Y$. In both cases, the world-volume theory has a
chiral $N=2$ global supersymmetry and the R-symmetry group is reduced
to $SU(2)$, which commutes with the holonomy group $SU(2)$ of
$K3$. The 6-dimensional space-time theory has an unbroken $N=2$
supersymmetry with central charge $(a,b=1,2$)
\be
\{Q^a,\overline{Q}^b\} = Z^{ab}.
\ee
As before, this symmetry is realized on the five-brane 
via the world-volume supersymmetry generators $\cQ^a$ and fermion 
zero-modes $S^a$. 

It is now straightforward to see that the two types of theories are
related via the $U$-duality (or $T$-duality) transformation that
interchanges the momentum and winding modes of the string theory on
the five-brane.  To see this we have to treat the two orbifolds
separately.  For the $K3\times S^1$ compactification, of the fluxes
introduced above for $T^5$, only the 8 $\Z_2$-invariant fluxes $q$,
$r_5$ and $s_{ij}$ with $i,j = 1,\ldots 4$ survive. In addition, there
are 16 extra fluxes 
\be
\label{extra}
s_I = \int_{S^1\times S_I^2} dU,
\ee
where the $S^2_I$ are the two-spheres surrounding the 16 fixed points
on $T^4/\Z_2$. The intersection form on the total set of fluxes has
signature (4,20), and the integers therefore label a 
vector\footnote{This `left-right' suffix will turn out to
correspond to the two chiral sectors of the space-time heterotic
string, and  should {\it not} be
confused with the label distinguishing left or right-moving 
string modes on the five-brane.}
$(p\LL,p\RR)$ in the lattice $\G^{4,20}$. 
In terms of the quadratic quantities $K_m,W_m$ the only 
non-vanishing component is given by
\be
K_5 = \half (p\LL^2 - p\RR^2).
\ee  
In the dual compactification on $T^4\times S^1/\Z_2$, on the other hand,
we are left with the fluxes $r_i$ and $s_{i5}$. These are complemented
by 16 extra fluxes $\tilde s_I$, which again combine with the other fluxes
into a vector $(\tilde p\LL,\tilde p\RR)\in \G^{4,20}$. 
In this case we find only
a non-zero contribution to
\be
W_5 = \half (\tilde p\LL^2 - \tilde p\RR^2).
\ee
As we will now show,
these results imply that after imposing the level matching constraints
$P^5\LL=P^5\RR=0$ and the BPS condition, only strings on the five-brane
with either pure momentum or pure winding number in the 5-direction 
will contribute in
BPS states. In particular, the T-duality map on this $S^1$ will interchange
the momentum and winding modes and thus the two compactifications. 

In both cases, the dependence of the central charge $Z$ on the integer fluxes
is parametrized by means of the action of $SO(4,20)$ on the
lattice $\G^{4,20}$. The central charge
only depends on the 4 `left-moving' components $p\LL$ (or $\tilde p\LL$)
via
\be 
Z^{ab} = p^i\LL\sigma_i^{ab}. 
\ee
BPS states again satisfy a condition of the form $(\e_aQ^a + \ebar_a
\Qbar^a)|\mbox{\small\sc BPS}\rangle =0$. 
In this case the central charge matrix $Z$ satisfies $Z^\dagger Z =
p\LL^2 {\bf 1}$, and the eigenvalue equation $Z^\dagger Z \e = m\bps^2
\e$ has therefore two independent solutions, with eigenvalue
\be 
m\bps = |p\LL|. 
\ee 
The resulting BPS multiplets are 16 dimensional, which equals the
dimension of the massless representations of the $N=(2,2)$ space-time
supersymmetry.

To obtain the multiplicities of the BPS states it is again necessary
to introduce a mode expansion of the low-energy string fields on the
world-volume. The modes are labeled by integral momentum $k=k^5$ and
winding number $w=w^5$ in the $S^1$ direction together with
a quantum number that labels the eigenmodes with eigenvalue $h$ of
the Laplacian on $K3$. The Hamiltonian takes the form
\be 
\cPo = \half p_i^2 + \half p\LL^2 + \half p\RR^2 + 
\sum_{k,w,h} \sqrt{k^2 + w^2 + h} \; N_{k,w,h} 
\ee 
Of the translation operators only the component $P^5$ in the
direction of the $S^1$ survives.
{}For the  $S^1 \times K3$ compactification it takes the form
\be
P^5 = \half p\LL^2 - \half p\RR^2 + \sum_{k,w,h} k N_{k,w,h}. 
\ee 
In a similar way as before, we deduce from combining the BPS
mass-shell condition $\cPo=\half p_i^2 + \half m^2\bps$ with the
constraint $\cP^5 = 0$, that charged BPS states can only oscillate in
the left-moving $S^1$ direction, {\it i.e.} opposite to $K^5$,
while the winding number must vanish $w=0$. The total number of
oscillators is constrained by the level matching condition
\be 
\half p\LL^2 - \half p\RR^2 + \sum_k k N_{k,0,0}=0.  
\ee 
In the dual compactification $T^4 \times S^1/\Z_2$ the only modification
is the interchange of the momentum label $k$ and the winding label $w$.
The total winding number of the string is related to the fluxes 
via the constraint $W^5=0$, which reduces for BPS states to
\be 
\half \tilde p\LL^2 - \half \tilde p\RR^2 + \sum_w w N_{0,w,0}=0.  
\ee 
This gives a concrete description of the BPS states in terms of the
underlying string modes. 

The number of oscillators that contribute to these
expressions is determined by the number of harmonic zero-modes with $h
= 0$ of the various fields on the K3 manifold.  There are 5 bosonic
oscillators $\aA_k^i$ with $i=1,\ldots 5$ corresponding to the
constant zero-modes of the coordinate fields $X^i$ and $Y$ on K3. The
19 anti-self-dual harmonic 2-forms on K3 lead to 19 additional
left-moving modes $\aA^i_k$ with $i= 6,\ldots, 24$.  So in the end we
are left with 24 left-moving bosonic oscillator modes $\aA_{k}^i$,
which can be recognized as the left-moving sector of the heterotic
string compactified on $T^4$ to 6 dimensions.\footnote{ The 3
self-dual two-forms on $K3$ and the chiral covariantly constant spinor
on $K3$ give rise to oscillator modes which are right-moving on $S^1$,
and form, together with the right-moving $X^i$ and $Y$ modes, the
chiral world-sheet content of the superstring.  These modes are
however all eliminated via the BPS restriction.}  Since also the
fluxes combine into a vector $(p\LL,p\RR)$ on $\G^{4,20}$, from this
point on the counting of BPS states exactly parallels that of the
heterotic string.

\newsection{Concluding Remarks}

In this paper we have presented a detailed analysis of the spectrum of
BPS states of the five-brane theory in eleven dimensions. We motivated
our proposed quantization procedure by using hints obtained from known
results about BPS states in six-dimensional string theory, such as the
symmetry under U-duality.  Although our formulation was not manifestly
Lorentz invariant in six dimensions, the final result for the spectrum
of BPS states is in fact invariant under the full Lorentz group
$SO(5,1)$.  This is a consequence of the fact that the BPS restriction
effectively reduces the five-brane dynamics to that of critical type
II superstring theory. Hence we can use the standard derivation to
show that our result for the complete BPS spectrum is indeed fully
covariant.

Our derivation should be compared with the analysis of D-brane states
\cite{polch}. In principle, it should be possible to make a concrete
identification between specific five-brane excitations and
configurations of D-branes and fundamental strings. Our construction
of the BPS spectrum in terms of a Fock space in section 6 indeed
matches the description of multiple D-brane configurations as given in
\cite{vafa6d}. Via this correspondence our results on the BPS spectrum
also give useful information about the bound states of strings and
D-branes \cite{witten-b}.

We have concentrated on compactifications to 6 dimensions. But it is
straightforward to extend our methods to compactifications to other
dimensions.  In particular, for dimensions higher than 6 one can
deduce the BPS spectra by simply putting some of the 16 charges equal
to zero. Thus, the results presented in this paper imply that at least
for compactifications to six dimensions and higher, all BPS states can
be obtained from the five-brane of M-theory. In particular, all the
states that one would naively associate with the two-brane have now
all become part of the spectrum of five-brane states.

This can be nicely illustrated in the specific example of the
compactification on $T^4\times S^1/\Z_2$ discussed in section 7 by
considering the decompactification limit for large volume of the
$T^4$. Here one obtains the ten-dimensional heterotic string states as
particular excitations of the five-brane. This should be compared to
the observations made in \cite{horava} that from the
eleven-dimensional point of view the heterotic string naturally arises
from the two-brane wrapped around the $S^1/\Z_2$. Our result however
suggest that one should be able to think of the two-brane as a
limiting configuration of the five-brane, in this case with the
world-brane topology of $K3\times S^1$. This could give an alternative
explanation of the $E_8\times E_8$ gauge symmetry.

To extend our formalism to dimensions five and lower, we have to
consider BPS states that are annihilated by 1/8 instead of 1/4 of the
space-time supercharges. This can be achieved by including vector
central charges (both in space-time and on the world-volume) that
modify the level matching conditions. In particular in five dimensions
this extends the flux sectors to all 27 charges, that transform in the
U-duality group $E_6(\Z)$ \cite{letter}. Hence the five-brane also
provides a unified description of all BPS states of 5 dimensional
string theory.  It would be interesting to see to which extent this
approach can be used to obtain the complete string BPS spectrum in
four dimensions.

\bigskip

{\noindent \sc Acknowledgements}

We would like to thank M. Becker, S. Ferrara, C. Kounnas, W. Lerche,
R.  Minasian, G. Moore, J. Polchinski, A. Strominger, C. Vafa, B. de
Wit and G. Zwart for interesting discussions and helpful comments.
This research is partly supported by a Pionier Fellowship of NWO, a
Fellowship of the Royal Dutch Academy of Sciences (K.N.A.W.), the
Packard Foundation and the A.P. Sloan Foundation.

\pagebreak

\renewcommand{\thesection}{A}
\renewcommand{\thesubsection}{A.\arabic{subsection}}
\addtocounter{section}{1}
\setcounter{equation}{0}
\setcounter{subsection}{0}
\setcounter{footnote}{0}
\begin{center}
{\large \sc Appendix}
\end{center}
\medskip

Our conventions are the following.  We denote the $SO(5)$
gamma-matrices as $\G_m{}^a{}_b$ with $m=1,\ldots,5$ and
$a,b=1,\ldots,b$. They satisfy
\be
\{\G_m,\G_n\} = 2\delta_{mn}, 
\ee 
and 
\be 
\G_m^\dagger = \G_m,\qquad
\G_m^T=\omega \Gamma_m \omega^T 
\ee 
with $\omega=-\omega^T$ the symplectic form that gives the isomorphism
$SO(5)\cong Sp(4)$. The other independent elements in the Clifford
algebra are of the form 
\be
\G_{mn}=\half[\G_m,\G_n],\qquad
\G_{mn}^\dagger = - \G_{mn} .
\ee 
We can use the matrices $\ww,\G_m,\G_{mn}$ to expand a general
matrix $Z$ that satisfies the reality condition 
\be 
Z^* = \omega Z,
\omega^T
\label{real}
\ee 
as 
\be
Z = a \ww + b^m\G_m + c^{mn}\G_{mn}. 
\ee 
Note that 
\be
Z^\dagger = a \ww + b^m\G_m - c^{mn}\G_{mn}, 
\ee 
so, if in addition $Z$ is hermitian, it is a linear combination of the
identity and $\G_m$. 

In the text, we will use the form $\omega_{ab}$ to lower and raise
indices of these matrices. Note that the matrices $\G_m^{ab}$ are now
antisymmetric. In particular we will use the notation
$\ww_{ab}=\omega_{ab}$.

A matrix $Z$ satisfying (\ref{real}) forms a 16-dimensional
Majorana-Weyl spinor representation of the group $SO(5,5)$. In fact
the $SO(5,5)$ gamma matrices can be written as 
\ba 
\G_m^L \is \G_m \otimes {\bf 1} \otimes i\sigma_2,\\ 
\G_m^R \is {\bf 1} \otimes \G_m \otimes \sigma_1, 
\ea 
and satisfy 
\be 
\{\G^L_m,\G^L_n\}=2\delta_{mn},\quad
\{\G^R_m,\G^R_n\}=-2\delta_{mn},\quad 
\{\G^L_m,\G^R_n\}=0.  
\ee 
In the chiral representation $S^\pm$ with $\G^{(11)}={\bf 1}\otimes
{\bf 1} \otimes \sigma_3=\pm 1$, the generators of the $SO(5)\otimes
SO(5)$ subgroup are given by $\G_{mn}\otimes {\bf 1},{\bf 1} \otimes
\G_{mn}$ whereas the off-diagonal generators are of the form $\pm \G_m
\otimes \G_n$.

\renewcommand{\Large}{\large}

\end{document}